\documentclass[onecolumn,preprintnumbers, superscriptaddress, reprint, longbibliography, pra]{revtex4-1}
\usepackage{amsmath}
\usepackage{stmaryrd}
\usepackage{txfonts}
\usepackage{amssymb}
\usepackage{mathrsfs}
\usepackage{graphicx}
\usepackage{dcolumn}
\usepackage{bm}
\usepackage{braket}
\usepackage{epsfig}
\usepackage{color}
\usepackage{ulem}
\usepackage{bibunits}

\usepackage[colorlinks=true, linkcolor=blue, citecolor=blue, urlcolor=blue]{hyperref}
\begin{document}

\title{Dimension transcendence and anomalous charge transport in magnets with moving multiple-$Q$ spin textures
}
\author{Ying Su}
\email{yingsu@lanl.gov}
\affiliation{Theoretical Division, T-4 and CNLS, Los Alamos National Laboratory, Los Alamos, New Mexico 87545, USA}
\author{Satoru Hayami}
\affiliation{Department of Physics, Hokkaido University, Sapporo 060-0810, Japan}
\author{Shi-Zeng Lin}
\email{szl@lanl.gov}
\affiliation{Theoretical Division, T-4 and CNLS, Los Alamos National Laboratory, Los Alamos, New Mexico 87545, USA}

\begin{abstract}
{Multiple-$Q$ spin textures, such as magnetic bubble and skyrmion lattices, have been observed in a large family of magnets.} These spin textures can be driven into motion by external stimuli. The motion of spin textures affects the electronic states. {Here we show that to describe correctly the electronic dynamics, the momentum space needs to be transcended to higher dimensions by including the ancillary dimensions associated with phason modes of the translational motion of the spin textures.} The electronic states have non-trivial topology characterized by the first and second Chern numbers in the high dimensional hybrid momentum space. This gives rise to an anomalous electric charge transport due to the motion of spin textures. By deforming the spin textures, a nonlinear response associated with the second Chern number can be induced {and results in the piezoelectricity}. {The charge transport is {derived from} the semi-classical equation of motion of electrons {that depends on the} Berry curvature in the hybrid momentum space.} {Our results suggest that the motion of multiple-$Q$ spin textures has significant effects on the electronic dynamics and provides a new platform to explore high dimensional topological physics.  }

\end{abstract}

\date{\today}
\maketitle

\section{Introduction}

Spatially localized spin textures with nontrivial topology, such as magnetic bubble and skyrmion, has been observed in several families of material systems including both metals and insulators, and heterostructures \cite{Muhlbauer2009,Yu2010a,Heinze2011,Seki2012}. These spin textures form a lattice, which gives rise to several Bragg peaks associated with the lattice symmetry in the spin structure factor \cite{Muhlbauer2009}.  For this reason, these crystallized spin textures are also named as multiple-$Q$ spin textures. The spin textures respond to various external stimuli, including electric current \cite{Jonietz2010,Yu2012,Schulz2012}, electric field \cite{White2012,Liu2013}, magnetic field \cite{Wang2015,zhang2018}, thermal gradient \cite{Kong2013,Lin2014PRL,Onose2010}, and strain \cite{PhysRevLett.115.267202,nii_uniaxial_2015,shibata_large_2015}, which renders them as very promising entities for applications \cite{Fert2013,nagaosa_topological_2013,wiesendanger_nanoscale_2016}. 

The \textit{static} spin texture has significant effects on electronic wave functions and affects the dynamics of conduction electrons \cite{NagaosaRMP2010,PhysRevB.98.235116}.  For instance, when electrons interact with a \textit{static} skyrmion, they experience an effective magnetic field produced by the noncoplanar spin texture. As a consequence, there is a topological Hall effect, which has been observed experimentally \cite{Neubauer2009,PhysRevLett.110.117202,PhysRevLett.112.186601}. The noncoplanar spin texture can open an energy gap in the electronic spectrum and stabilize a Chern insulator \cite{PhysRevLett.101.156402,batista_frustration_2016,sorn_tunable_2018,PhysRevB.98.235116}. 
The spin texture can be driven into motion by  magnetic or electric field gradients \cite{Liu2013,Wang2015,zhang2018}, thermal gradients \cite{Kong2013,Lin2014PRL,Onose2010}, and spin-polarized currents \cite{Zang2011,Lin2013PRL,Iwasaki2013}. How the \textit{motion} of spin textures affects the electronic dynamics is an important question. 
{In ferromagnetic metals, the interaction between conduction electrons and moving spin textures can generate an emergent electric field that results in a spin motive force affecting the  electronic dynamics \cite{Barnes2007,Yang2009,Hals2015,Schulz2012}. 
Nevertheless, how it works for a gapped system is not well understood.}

In this {paper}, we reveal that the description of the dynamics of conduction electrons coupled to the moving multiple-$Q$ spin textures requires a transcendent high dimensional momentum space. The high dimensional momentum space is spanned by the physical dimensions and ancillary dimensions associated with phason modes of the translational motion of spin textures. The electronic states can have nontrivial topology characterized by the first and second Chern numbers in the hybrid momentum space when the energy spectrum is gapped. Due to the nontrivial topology, there is an anomalous charge transport in magnets with \textit{moving} multiple-$Q$ spin textures. {To explain the anomalous charge transport, we develop a general semi-classical theory to study the response of electronic states to the motion and deformation of spin textures.} Our results demonstrate that  the motion of multiple-$Q$ spin textures has profound consequences on the electronic wave functions and dynamics.



\section{Model and dimension transcendence}

The interaction between conduction electrons and spin textures can be described by the Hamiltonian
\begin{equation}
\mathcal{H} = -t\sum_{\langle i,j \rangle} c_i^\dagger c_j - J \sum_i  c_i^\dagger \bm{S}_i\cdot\bm{\sigma} c_i - B\sum_i c_i^\dagger \sigma_z c_i,
\label{H}
\end{equation}
where $c_i=(c_{i,\uparrow},c_{i,\downarrow})^\top$ is the annihilation operator of conduction electrons with spin $\uparrow$ and $\downarrow$ at the $i$th site, $\langle i,j \rangle$ denotes two nearest-neighbor sites on a 2D square lattice, $\bm{\sigma}$ represents the vector of Pauli matrices, and $\bm{S}_i$ encodes the spin texture. Here we assume that the spin texture $\bm{S}_i$ is stabilized by other stronger interactions and is not affected by the conduction electrons. The spin configuration of the multiple-$Q$ spin textures can be generally described by
\begin{equation}
\bm{S_i}=\sum_\nu {G}_\nu^\mu(\bm{Q}_\nu \cdot\bm{r}_i+\phi_\nu) \hat{\bm{e}}_\mu  ,  
\end{equation}
where ${G}_\nu^\mu(\phi_\nu)={G}_\nu^\mu(\phi_\nu+2\pi)$ and the Einstein summation convention is employed  hereafter. The translational motion of the spin texture, $\bm{r}_i\rightarrow\bm{r}_i+ \bm{v}
\tau$, is depicted by the phase $\phi_\nu=\bm{Q}_\nu\cdot\bm{v}
\tau$, where $\bm{v}$ is velocity and $\tau$ is time.

The coupling of conduction electrons to multiple-$Q$ spin textures yields a magnetic superlattice. We denote the crystal momentum of the magnetic superlattice as $\bm{k}$. The Hamiltonian in the momentum space is also a periodic function of $\phi_\nu$ besides $\bm{k}$, i.e. $H(\bm{k},\bm{\phi})$ with $\bm{\phi}=(\phi_1,\phi_2,\cdots)$. Therefore, when the spin texture is moving, the electronic dynamics is depicted in the effective momentum space spanned by $\tilde{\bm{k}}=\bm{k}\oplus \bm{\phi} {=(k_x,k_y,\phi_1,\phi_2,\cdots)}$, a hybridization of crystal momenta and phason modes of the translational motion of spin textures. {The motion of the spin textures is slow compared to electronic dynamics. For instance, for a spin texture with period $1$ nm moving at a velocity $1\ \mathrm{m/s}$, $\phi_\nu$ changes at a rate of 1 GHz. As a good approximation, the {electronic} spectrum {$E_{ n}({\tilde{\bm{k}}})$}  and the corresponding eigenstate {$\ket{\psi_{n\tilde{\bm{k}}}}$} evolve adiabatically with $\phi_\nu$.}


\section{Nonlinear response}
The semi-classical equation of motion for an electronic wave packet with  the dispersion $E_n(\tilde{\bm{k}})$ and {center of mass $\bm{r}_{n}$} follows
\begin{equation}\label{eq3}
\dot{r}^{\mu}_{n}  =\frac{\partial E_n  (\tilde{\bm{k}}) }{\hbar \partial \tilde{k}_{\mu}}-\dot{\tilde{k}}_{\nu }  \Omega_n^{\mu \nu }(\tilde{\bm{k}}), 
\end{equation}
{where $\Omega_n^{\mu \nu }=i\partial_{\tilde{k}_\mu} \braket{u_{n\tilde{\bm{k}}}| \partial_{\tilde{k}_\nu}u_{n\tilde{\bm{k}}}} - i\partial_{\tilde{k}_\nu} \braket{u_{n\tilde{\bm{k}}}| \partial_{\tilde{k}_\mu}u_{n\tilde{\bm{k}}}}$ is the Berry curvature and $\ket{{u_{n\tilde{\bm{k}}}}}$ is the periodic part of the Bloch state $\ket{\psi_{n\tilde{k}}}$. 
Eq. (\ref{eq3}) is meaningful only when $\mu=x$ or $y$ since the center of mass $\bm{r}_n$ is confined in the real space.} Here the first term is the group velocity and the last term is the anomalous velocity due to nonzero Berry curvatures \cite{Xiao2010}.
In the absence of external electromagnetic fields, $\dot{\tilde{k}}_\nu=\dot{k}_\nu=0$ for $\nu=\{x,y\}$. 
However, the translational motion of spin textures described by $\phi_\nu=\omega_\nu \tau$ ensures a nonzero $\dot{\tilde{k}}_\nu=\dot{\phi}_\nu=\omega_\nu$ for $\nu=\{1,2,\cdots\}$.
{In addition to the conventional anomalous velocity $-\dot{k}_\nu\Omega_n^{\mu \nu }(\tilde{\bm{k}})$ \cite{Xiao2010}, there is a new contribution $-\dot{\phi}_\nu\Omega_n^{\mu \nu }(\tilde{\bm{k}})$  that can be obtained in the semi-classical theory (see Appendix \ref{AA}-\ref{AC}). The Berry curvature in the new contribution is defined in the hybrid momentum   space consist of the crystal momentum $k_\mu$ and phason mode $\phi_\nu$. 
The semi-classical equation of motion is derived in the leading order.  The higher order corrections do not change the form of equation of motion but modify the dispersion relation and Berry curvature due to interband coupling (see Appendix \ref{AB}). Since the higher order corrections do not change our finally results (proved in Appendix \ref{AD}), the modifications of dispersion relation and Berry curvature are not considered here.}
The corresponding current density due to the translational motion of spin textures is
\begin{equation}\label{j0}
     j^{\mu }_0=e\sum_n\int \frac{d^2 {k}} {4 \pi^2} f\left(E_n-E_F\right)\left(\frac{\partial E_n }{\hbar \partial k_{\mu }}-\omega _{\nu }  \Omega_n^{\mu \nu }\right),
\end{equation}
where $f\left(E_n-E_F\right)$ is the Fermi distribution function and $E_F$ is the Fermi energy. When the energy spectrum is gapped in the hybrid momentum space and the Fermi energy $E_F$ is in the gap, the first term in Eq. \eqref{j0} vanishes. The transported charge over one period $\Delta \tau ={2 \pi }/{|\omega _{\nu }|}$ at zero temperature is 
\begin{equation}\label{q0}
    q^{\mu }_0=-\frac{eN^\mu}{|\omega_\nu|}\sum_{E_n\le E_F} \int_0^{2\pi} d\phi _{\nu }\int \frac{d^2k}{4\pi^2}\omega_\nu\Omega_n^{\mu \nu }=\frac{eN^\mu \omega_\nu C_1^{\mu \nu }}{|\omega_\nu|},
\end{equation}
\begin{equation}\label{C1}
C_1^{\mu \nu}=-\frac{1}{2\pi}\sum_{E_n\le E_F} \int dk_\mu d\phi_\nu\Omega_n^{\mu \nu },
\end{equation}
where $C_1^{\mu \nu}$ is the first Chern number defined on the $\tilde{k}_\mu\tilde{k}_\nu$ plane and is independent of the other momenta if the energy gap retains open in the entire hybrid momentum space, and $N^\mu$ is the number of unit cells in the cross section perpendicular to the $\mu$ direction. Therefore the translational motion of the spin textures can result in quantized charge transport in magnetic insulators with $C_1^{\mu \nu }\neq 0$, which is the topological pumping.


The response function Eq. (\ref{j0}) becomes nonlinear when the dynamics of $\phi_\nu$ couples with ${\dot{\bm{r}}_n}$. This can be achieved by deforming the spin texture, $\bm{Q}_{\nu}\rightarrow \bm{Q}_{\nu}+\bm{Q}'_{\nu}$. Therefore, $\phi_\nu=\omega_
\nu\tau +\bm{Q}'_{\nu}\cdot \bm{r}_{n}$, and its dynamics follows 
\begin{equation}\label{k}
    \dot{\tilde{k}}_\nu=\dot{\phi }_{\nu }=\omega _{\nu }+{Q}_{\nu\mu}' \dot{{r}}^\mu_{n},
\end{equation}
{that depends on the velocity of wave packet (see Appendix \ref{AA}).}
Substituting Eq. (\ref{k}) into Eq. (\ref{eq3}) and retaining the terms up to the second order in $\omega_\nu$ and $Q_{\nu\mu}'$, we obtain
\begin{equation}\label{r2}
\dot{r}^{\mu }_{n}=\frac{\partial E_n  }{\hbar \partial k_{\mu }}-\omega _{\nu }  \Omega_n^{\mu \nu }-\left(\frac{\partial E_n}{\hbar \partial k_{\gamma }}-\omega _{\delta }  \Omega_n^{\gamma \delta }-\frac{\partial E_n }{\hbar \partial k_{\eta}}Q_{ \delta\eta}' \Omega_n^{\gamma \delta }\right)Q_{\nu \gamma }' \Omega_n^{\mu \nu },
\end{equation}
where the first two terms correspond to the linear response in Eq. (\ref{j0}). The last term in Eq. (\ref{r2}) originated from the deformation of spin textures induces a nonlinear response 
\begin{equation}\label{j}
j^\mu = j_0^\mu + e\sum_n\int \frac{d^2k}{{8}\pi^2} f(E_n-E_F)F_n^{\mu\nu\gamma\delta} \omega_\nu Q_{\gamma\delta}',
\end{equation} 
where  $F_n^{\mu\nu\gamma\delta}= \Omega_n^{\mu \nu } \Omega_n^{\gamma \delta}+\Omega_n^{\mu \gamma } \Omega_n^{\delta \nu}+\Omega_n^{\mu \delta } \Omega_n^{\nu\gamma }$ {(see Appendix \ref{AD})}.
In this case, the transported charge over $\Delta \tau$ at zero temperature is 
\begin{equation}\label{q}
\begin{split}
q^\mu&=q_0^\mu + \frac{eL^\mu}{|\omega_\nu|}\int_0^{2\pi} d\phi_\nu\sum_{E_n\le E_F}\int \frac{d^2k}{{8}\pi^2} F_n^{\mu\nu\gamma\delta}  \omega_\nu Q_{\gamma\delta}'   \\
&\thickapprox q_0^\mu  + \frac{eL^\mu C_2^{\mu \nu \gamma \delta} \omega_\nu Q_{\gamma\delta}'}{{4}\pi |\omega_\nu|} 
\end{split}
\end{equation}
where $L^\mu$ is the size of the cross section perpendicular to the $\mu$ direction and
\begin{equation}\label{C2}
C_2^{\mu \nu \gamma \delta}=\frac{1}{4\pi^2}\sum_{E_n\le E_F}\int dk_\mu  d\phi_\nu dk_\gamma d\phi_\delta F_n^{\mu\nu\gamma\delta},
\end{equation}
is the second Chern number \cite{Yang_Generalization_1978} defined on the $\tilde{k}_\mu\tilde{k}_\nu\tilde{k}_\gamma\tilde{k}_\delta$ hypersurface of the hybrid momentum space. Here  we make the approximation that $F_n^{\mu\nu\gamma\delta}\thickapprox \int d\phi_\delta F_n^{\mu\nu\gamma\delta} /2\pi$. The approximation is valid when the Berry curvatures are weakly dispersive as a function of $\phi_\delta$. This is true when the spin texture is incommensurate with the lattice or the period of spin texture is much larger than the lattice constant. 
In our definition, both $C_1^{\mu\nu}$ and $C_2^{\mu \nu \gamma \delta}$ are antisymmetric under the permutation of the indices   because  $\Omega_n^{\mu \nu }=-\Omega_n^{\nu \mu }$. Equation (\ref{q}) demonstrates that, in addition to the contribution from the first Chern number, the charge transport has the contribution from the second Chern number when the spin texture is deformed.

\begin{figure}
  \begin{center}
  \includegraphics[width=8.5cm]{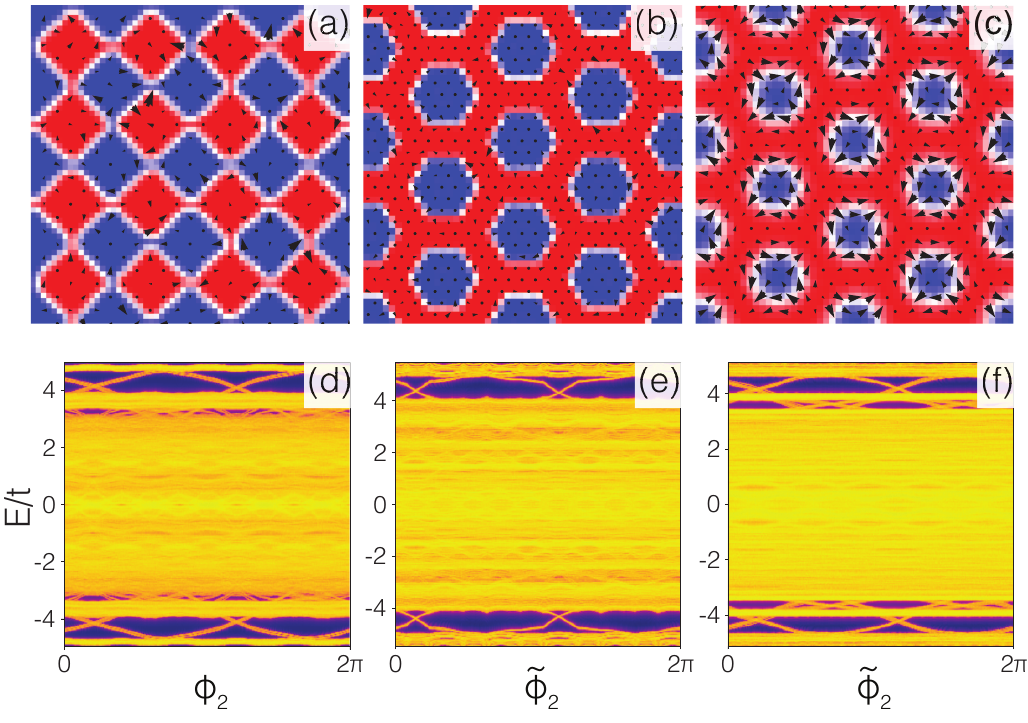}
  \end{center}
\caption{Spin textures obtained from the Monte Carlo simulation. {(a) The double-$Q$ collinear magnetic bubble lattice. (b) The triple-$Q$ collinear magnetic bubble lattice. (c) The triple-$Q$ skyrmion lattice. Here the arrows represent the in-plane components and the background color (with red for positive and blue for negative) stands for the $z$ component of local spins. (d)-(f) Electronic spectral functions as a function of $\phi_2$ ($\tilde{\phi}_2$) for conduction electrons coupled to the three spin textures (a)-(c), respectively.} Here the purple (yellow) denotes low (high) density of states. }  
  \label{fig1}
\end{figure}

To substantiate the anomalous charge transport due to the translational motion of the spin textures from the semi-classical analysis above, we study the charge transport in magnets with several typical multiple-$Q$ spin textures obtained from the Monte Carlo simulation (see Appendix \ref{AF}). Here we consider three kinds of spin textures: double-$Q$ and triple-$Q$ collinear magnetic bubble lattices {where spins are nearly parallel or antiparallel due to the strong easy axis anisotropy}, and triple-$Q$ skyrmion lattice as shown in Figs. \ref{fig1}(a)-\ref{fig1}(c), respectively. For conduction electrons hopping on a square lattice (whose lattice constant is set to unity), the $\bm{Q}_\nu$ vectors for the double-$Q$ magnetic bubble lattice are $\bm{Q}_1=(2\pi/5,0)$ and $\bm{Q}_2=(0,2\pi/5)$, while for the triple-$Q$ magnetic bubble and skyrmion lattices are $\bm{Q}_1=(-2\pi/5,2\sqrt{3}\pi/5)$, $\bm{Q}_2=(-2\pi/5,-2\sqrt{3}\pi/5)$, and $\bm{Q}_3=(4\pi/5,0)$.  For the triple-$Q$ states, because $\sum_{\nu=1}^3\phi_\nu=0$ for a translational motion, only two of $\phi_\nu$ are independent of each other.  

\section{Topological property} 
To identify the topology of electronic states in magnets with the multiple-$Q$ spin textures, we first study their electronic spectra in the hybrid momentum space. For the double-$Q$ magnetic bubble lattice, the translational motion of the spin texture is parameterized by $\phi_\nu=\omega_\nu\tau$. The energy spectrum as a function of $\phi_2$ (with $\phi_1=0$) is shown in Figs. \ref{fig1}(d) for the double-$Q$ spin texture with $J=1.5t$ and $B=0$. Here we use the periodic boundary condition along the $x$ direction and the open boundary condition along the $y$ direction. Apparently, there are topological edge states in the bulk energy gaps indicating the system is topologically nontrivial. 
For the triple-$Q$ spin textures, to make them commensurate with the lattice, we compress the lattice constant along the $y$ direction to $1/\sqrt{3}$. In this case, because $\sum_{\nu=1}^3\phi_\nu=0$, we define $\tilde{\phi}_1=-(\phi_1+\phi_2)/2=\phi_3/2$ and $\tilde{\phi}_2=(\phi_1-\phi_2)/2$ that are independent. The change of $\tilde{\phi}_1$ and $\tilde{\phi}_2$ corresponds to the shift of triple-$Q$ spin textures along the $x$ and $y$ directions, respectively. Thus we can parameterize the translational motion of the triple-$Q$ spin textures by $\tilde{\phi}_\nu=\omega_\nu\tau$. The energy spectra as a function of $\tilde{\phi}_2$ (with $\tilde{\phi}_1=0$) for the triple-$Q$ magnetic bubble and skyrmion lattices with $J=1.5t$ and $B=-1.2t$ are displayed in Figs. \ref{fig1}(e) and \ref{fig1}(f), respectively. The topological edge states and  bulk energy gaps persist for the triple-$Q$ spin textures.

To characterize the nontrivial band topology, we calculate the Chern numbers of the system with an efficient algorithm by using the $U(1)$ link variable \cite{Fukui_Chern_2005, Mochol_Efficient_2018}. For commensurate spin textures, the system has translational symmetry and we can describe it in a high dimensional hybrid momentum space spanned by the generalized crystal momentum $\tilde{\bm{k}}=(k_x,k_y, \phi_1, \phi_2)$ for the double-$Q$ spin texture and $\tilde{\bm{k}}=(k_x,k_y,\tilde{\phi}_1,\tilde{\phi}_2)$ for the triple-$Q$ spin textures.  Even if the spin texture is incommensurate with the lattice, $k_x$ and $k_y$ can still be introduced by using the twisted boundary condition \cite{Sheng_Spin_2005}. Consequently, the Chern numbers in Eqs. (\ref{C1}) and (\ref{C2})  are well defined on the compact manifold. As an example, we consider the Fermi energy in the bottom
bulk energy gaps in Figs. \ref{fig1}(d)-\ref{fig1}(f). In the 4D hybrid momentum space, there are six first Chern numbers:  $C_1^{x1}=C_1^{y2}=2$ and $C_1^{xy}=C_1^{x2}=C_1^{y1}=C_1^{12}=0$, and one second Chern number: $C_2^{x1y2}=-2$ for the double magnetic bubble lattice. For the triple-$Q$ magnetic bubble and skyrmion lattices, they have the same Chern numbers: $C_1^{x1}=C_1^{y2}=2$, $C_1^{xy}=C_1^{x2}=C_1^{y1}=C_1^{12}=0$, and $C_2^{x1y2}=-4$. {Interestingly, our results demonstrate an insulator that is trivial in the crystal momentum space, can be topologically nontrivial in the hybrid momentum space. }

\begin{figure*}[t]
  \begin{center}
  \includegraphics[width=15 cm]{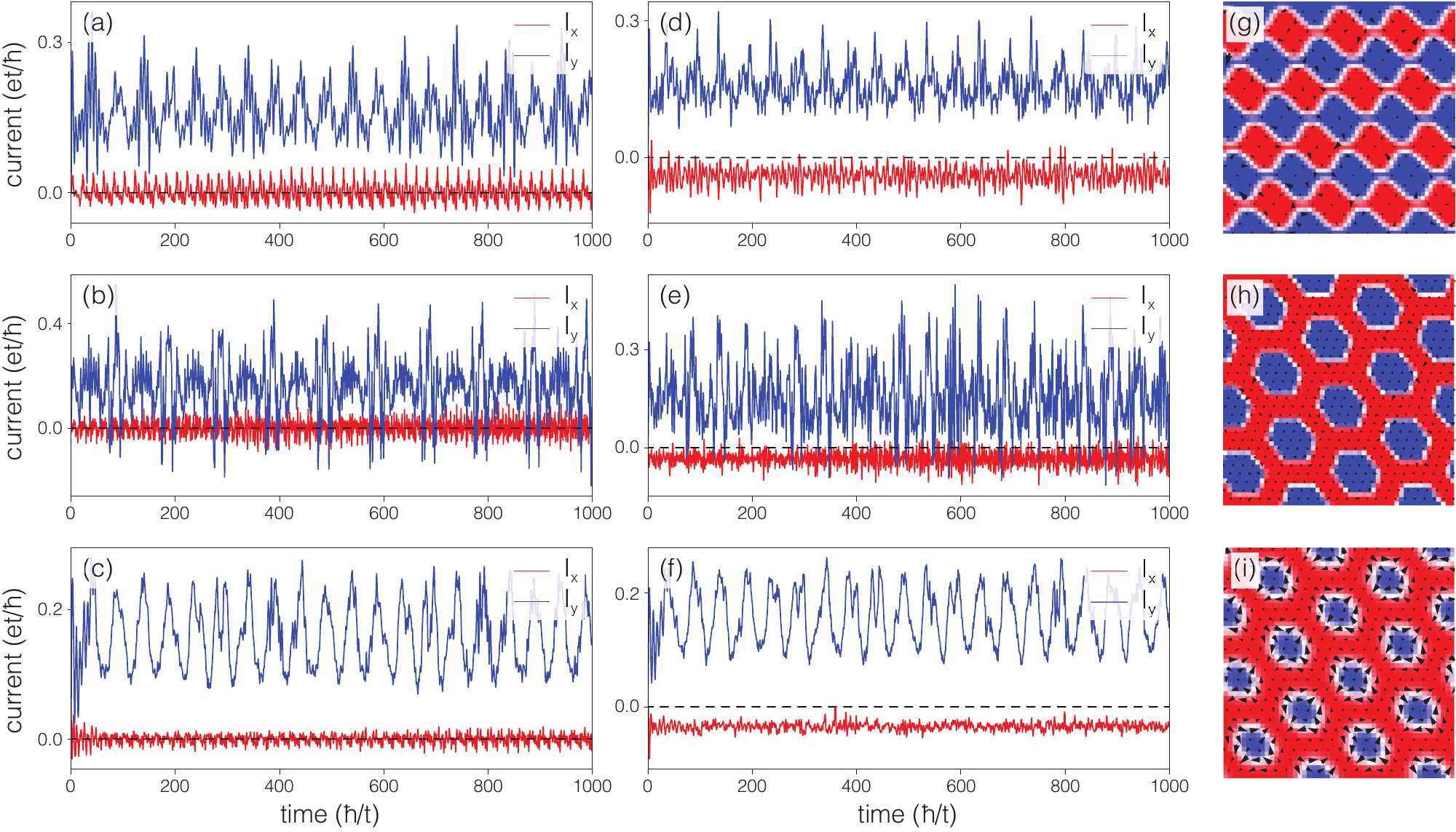}
  \end{center}
\caption{Currents along the $x$ and $y$ directions {induced by} the translational motion of spin textures. (a) Currents for the double-$Q$ magnetic bubble lattice in Fig. \ref{fig1}(a), (b) currents for the triple-$Q$ magnetic bubble lattice in Fig. \ref{fig1}(b), and (c) currents for the triple-$Q$ skyrmion lattice in Fig. \ref{fig1}(c). (d)-(f) Currents for the three spin textures under a shear deformation. (g)-(i) Sheared spin textures are obtained by applying a shear strain on the spin textures in Fig. \ref{fig1}(a)-(c). } 
  \label{fig4}
\end{figure*}

\section{Anomalous charge transport} 
The translational motion of spin textures can induce a quantized charge transport according to Eqs. (\ref{q0}) and (\ref{C1}). Here we consider the spin textures shifting along the $y$ direction on a $40\times40$ lattice (containing $8\times8$ supercells) with $\omega_1=0$ and $\omega_2=2\pi t/100\hbar$, and we calculate the electric currents of the system by solving the time-dependent Schr\"{o}dinger equation (see Appendix \ref{AG}). We focus on the Fermi energy in the bottom bulk energy gaps in Figs. \ref{fig1}(d)-\ref{fig1}(f). For the double-$Q$ magnetic bubble lattice, the currents along the $x$ and $y$ directions over ten periods are shown in Fig. \ref{fig4}(a). By integrating the currents over time, the average transported charges in one period are $q^x=0.01e$ and $q^y=15.95e$. According to Eq. (\ref{q0}),  $q^x=eN^x\omega_2 C^{x2}/|\omega_2|=0$ and $q^y=eN^y\omega_2 C^{y2}/|\omega_2|=16e$. Apparently, the number obtained by simulation shows a good agreement with that obtained from the semi-classical equation. For the triple-$Q$ magnetic bubble lattice, we have $q^x=0.14e$ and $q^y=16.04e$ from the currents in Fig. \ref{fig4}(b), while for the triple-$Q$ skyrmion lattice, we have $q^x=0.00e$ and $q^y=15.96e$ from the currents in Fig. \ref{fig4}(c). The numerical results for the triple-$Q$ spin textures are also consistent with $q^x=0$ and $q^y=16e$ obtained from Eq. (\ref{q0}). 

The nonlinear response can be realized by deforming the spin texture as shown in Eqs. (\ref{k}) and (\ref{j}). For the double-$Q$ magnetic bubble lattice, we deform the spin texture by setting $\phi_1=\bm{Q}_1'\cdot\bm{r}$ with $\bm{Q}_1'=(0,2\pi/20)$. This induces a shear strain $\gamma_{xy}=-Q_{1y}'/Q_{1x}=-0.25$ on the spin texture that results in $x\rightarrow x +\gamma_{xy}y$, as shown in Fig. \ref{fig4}(g). The currents due to the translational motion and shear strain of the spin texture are displayed in Fig. \ref{fig4}(d). In this case, the transported charges in one period are $q^x=-3.92e$ and $q^y=15.89e$ from the currents in Fig. \ref{fig4}(d). Meanwhile, according to the Eq. (\ref{q}), we have $q^x=eL^xC_2^{x21y}\omega_2Q_{1y}'/2\pi |\omega_2|=-4e$ where $C_2^{x21y}=C_2^{x1y2}=-2$, and $q^y=16e$ as before. The results from numerical simulation and semi-classical analysis are consistent with each other. Therefore, we demonstrate that, besides the longitudinal charge transport due to the linear response, there is a transverse charge transport caused by the nonlinear response of the deformed spin texture. Moreover, one can probe the Chern numbers experimentally from the quantized charge transport. For the triple-$Q$ spin textures, the shear strain can also be induced by taking $\tilde{\phi}_1=\bm{Q}_1'\cdot\bm{r}$, where we set $\bm{Q}_1'=(0,2\sqrt{3}\pi/40)$ and $\gamma_{xy}=Q_{1y}'/Q_{1x}=-0.125\sqrt{3}$ in Figs. \ref{fig4}(h) and \ref{fig4}(i). The currents of the triple-$Q$ magnetic bubble lattice are shown in Fig. \ref{fig4}(e) and the transported charges in one periodic are $q^x=-3.43e$ and $q^y=15.60e$. The currents of the triple-$Q$ skyrmion lattice are shown in Fig. \ref{fig4}(f) and the transported charges are $q^x=-3.48e$ and $q^y=15.92e$. Equation (\ref{q}) yields $q^x=-4e$ and $q^y=16e$ for the triple-$Q$ spin textures. 
{In this case, the numerical results are less quantized.  One possible reason is due to the fact that the spin textures from Monte Carlo simulation are not exactly periodic. As a comparison, we also calculate the current for the double-$Q$ and triple-$Q$ spin texture ansatz described by the linear superposition of cosine functions, and the transported charge for the ansatz are  perfectly quantized (see Appendix \ref{AE}). }

{When the deformed spin texture is commensurate with the lattice, one may wonder whether the anomalous charge transport can be fully characterized by the first Chern numbers of the deformed system. To answer this question, we take the deformed double-$Q$ magnetic bubble lattice studied above as an example.  Under the deformation induced by  $\bm{Q}_1'=(0,2\pi/20)$, the new unit cell is enlarged to contain $5\times20$ lattice sites (fourth larger than the undeformed unit cell that is $5\times 5$). Therefore on the $40\times40$ lattice, $N_x=2$ and $N_y=8$. Because the lattice constant along the $y$ direction is increased by 4 times, the period of motion of the spin texture is also increased by 4 times (since  $\omega_2$ is fixed).  By performing exactly the same calculation, we find the relevant first Chern numbers of the deformed system become $C_1^{x2}=-8$ and $C^{y2}_1=8$. Plugging these into Eq. (\ref{q0}), we get $q^x = -16e$ and $q^y=64e$ in one period of motion. To compare with the previous results, we need to divide the new $q^x$ and $q^y$ by 4 to count the transported charges in the same period of time as  the undeformed case.   Thus, the two independent approaches yield the same result. As shown in this example, even a small deformation of spin textures can increase the size of unit cell significantly. With the periodic modulation extended to a larger length scale, there are more energy bands due to the band folding, accompanying complicated band gap closing and opening in this process, where the topological phase transitions are  possible. Therefore, without the nonlinear response function associated with the second Chern number, one needs to calculate different first Chern numbers for different deformation to understand the charge transport. This calculation turns out to be impractical for small deformation that results in huge unit cells. In this sense, the nonlinear response theory can make life much easier since one only needs to know the first and second Chern numbers of the undeformed system and these Chern numbers are usually relatively easy to calculate.}

\section{Linear piezoelectricity.}
\begin{figure}[t]
  \begin{center}
  \includegraphics[width=8.5 cm]{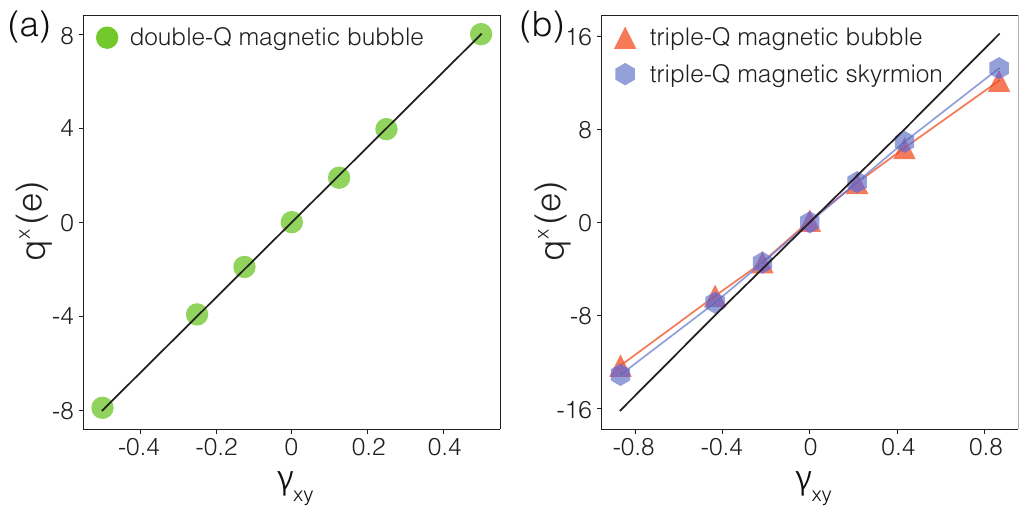}
  \end{center}
\caption{Linear piezoelectricity induced by the deformation of spin textures. (a) $q^x$ as a function of {the} {deformation} $\gamma_{xy}$ for the double-$Q$ magnetic bubble lattice, and (b) for the triple-$Q$ magnetic bubble and skyrmion lattices. The black lines are from the response function Eq. (\ref{q}).} 
  \label{figs3}
\end{figure}

As we have shown above, the deformation of spin textures is essential to generate the nonlinear response. 
{The deformation of magnetic skyrmion lattices has been realized in a strained crystal \cite{shibata_large_2015} or by applying an electric field \cite{White2014}. Very recently, the deformation of the moving magnetic skyrmion lattice under electric current flow has also been observed \cite{Okuyama2019}.} According to Eq. (\ref{j}), the current depends linearly on the deformed $\bm{Q}_v'$ vector, i.e. $j^\mu\propto Q_{\gamma\delta}'$. 
For instance, when the spin texture is deformed by an effective shear strain  $\gamma_{xy}\propto Q_{1y}'$ in our numerical simulation above, the linear piezoelectric current is $j^x\propto \gamma_{xy}$. To verify the linear piezoelectricity, we calculate  $q^x$ as a function of $\gamma_{xy}$ and we expect $q^x\propto \gamma_{xy}$. The numerical results are displayed in Fig. \ref{figs3}. For the finite system size ($40\times 40$ lattice sites) considered in our numerical calculations, the shear strain are limited to a few discrete values in order to retain the periodic boundary condition. For much larger systems, the shear strain can be tuned continuously. We show $q^x$ as a function of $\gamma_{xy}$ for the double-$Q$ magnetic bubble lattice in Fig. \ref{figs3}(a), and for the triple-$Q$ magnetic bubble and skyrmion lattices in Fig. \ref{figs3}(b). Here the symbols are from the numerical calculation and the black lines are from the Eq. (\ref{q}). Apparently, the $q^x$ for the double-$Q$ spin texture collapses on the black line and shows a good linear dependence on $\gamma_{xy}$. 
The $q^x$ for the triple-$Q$ spin textures are also linear in $\gamma_{xy}$ but deviate from the black line due to poor quantization of charge transport as reasoned above. 
Therefore, the linear piezoelectricity is verified numerically and works for moderate strains that preserve the intrinsic topology. Our results suggest a new mechanism of piezoelectricity as a consequence of the nontrivial topology.

\section{Discussion}

{The nonlinear response due to the translational motion and deformation of the spin textures is associated with the second Chern number.} The quantized charge transport requires the Fermi energy in the bulk energy gap. When this is not the case, Eq. (\ref{j}) is still valid {but the transported charge is not quantized}. Therefore, the anomalous charge transport generated by the motion of spin textures occurs both in insulating and metallic magnets.
 {In our system, the electronic dynamics is governed by the nontrivial topology of electronic structure in the hybrid momentum space spanned by the crystal momentum and phason mode of the spin texture. This is different from the emergent electrodynamics induced by the motion of non-collinear spin textures according to the generalized Faraday’s law \cite{Barnes2007,Yang2009,Schulz2012} or magnetization
dynamics as a reciprocal process of the spin transfer torque \cite{Hals2015}. Evidently, the anomalous charge transport in the collinear spin textures and in magnetic insulators can not be captured by the emergent electrodynamics.} 
Our predictions can be tested in B20 chiral magnets \cite{Muhlbauer2009,Yu2010a}, $f$-electron materials \cite{garnier_magnetic_1998,PhysRevLett.120.097201} and magnetic heterostructures \cite{Jiang17072015,jiang_direct_2017,litzius_skyrmion_2017}, where multiple-$Q$ spin textures have been observed in experiments. 
The quantized charge transport can be possibly realized in a heterostructure with the multiple-$Q$ spin texture formed in a magnetic insulator layer or at the interface through the proximity effect \cite{shao2019,ding2019,Li2014}.  In materials, the multiple-$Q$ spin textures are pinned by defects. To drive the spin texture into motion, a large driving force is needed in order to overcome the pinning energy. The pinning-depinning transition provides a clear way to distinguish the anomalous charge transport from other contributions.


In summary, we reveal that the motion of spin textures has significant effects on the dynamics of conduction electrons. The semi-classical equation of motion of the electronic wave packet is described in a high dimensional hybrid momentum space. The hybrid momentum space is constituted of the physical dimensions and ancillary dimensions associated with the phason modes of the translational motion of the spin textures. The electronic states can be topologically nontrivial when the energy spectrum is gapped, and the nontrivial topology is characterized by the first and second Chern numbers in the hybrid momentum space. The nontrivial topology results in an anomalous charge transport. The charge transport is quantized and can be used to extract the Chern numbers experimentally. {Our theory presents an entirely new mechanism for electric current generation induced by the motion of multiple-Q spin texture in magnetic insulators.} Our results can be extended to 3D magnets with multiple-$Q$ spin textures where even higher dimensional topological physics can be realized.

\begin{acknowledgements}
We would like to thank Joel Moore, {Chih-Chun Chien, and Cristian Batista} for insightful discussions. Computer resources for numerical calculations were supported by the Institutional Computing Program at LANL. This work was carried out under the auspices of the U.S. DOE NNSA under contract No. 89233218CNA000001 through the LDRD Program, and was supported by the Center for Nonlinear Studies at LANL (Y. S. and S.Z.L). S.H. is supported by the JSPS KAKENHI Grants No. JP18K13488.
\end{acknowledgements}

\appendix
\section{Semi-classical equation of motion}\label{AA}
Here we consider a very general model of free electrons coupled to a periodic modulation potential described by
\begin{equation}
H_0 = \frac{\bm{p}^2}{2m} + \sum_\nu V_\nu(\bm{Q}_\nu \cdot \bm{r}),
\label{H0}
\end{equation}
and we study how the electron dynamics is affected by the motion and deformation of the potential. {The potential is the linear superposition of a set of periodic functions $V_\nu(\bm{Q}_\nu \cdot \bm{r})$ with the modulating vector $\bm{Q}_\nu$.} The perturbation can be induced by asking $V_\nu(\bm{Q}_\nu \cdot \bm{r}) \rightarrow V_\nu(\bm{Q}_\nu \cdot \bm{r} + \phi_\nu)$ where 
\begin{equation}
\phi_\nu (\bm{r},\tau)= \omega_\nu \tau  + \bm{Q}_\nu'\cdot\bm{r}.
\end{equation}
Here the first term shifts the potential and the second term deforms the potential. Apparently, the multiple-$Q$ spin texture model belongs to the category depicted by Eq. (\ref{H0}). In the semi-classical approach, the electronic state is described by a wave packet that is highly localized at the center of mass $\bm{r}_c$ and $\bm{k}^c$ in the phase space. Due to the localized nature, when the perturbation is smooth compared to the length scale spread by the wave
packet, the approximate Hamiltonian for the wave packet can be obtained by linearizing the perturbation as $\phi_\nu(\bm{r},\tau) \approx \phi_\nu(\bm{r}_c,\tau)$ \cite{Xiao2010}. 
The leading order local Hamiltonian is 
\begin{equation}
H_c=\frac{\bm{p}^2}{2m} +  \sum_\nu V_\nu(\bm{Q}_\nu \cdot \bm{r} + \phi_\nu (\bm{r}_c, \tau))
\end{equation} 
and the first order correction is $H_1 = \partial_{\bm{r}_c} H_c \cdot (\bm{r}-\bm{r}_c)$. Here we have 
\begin{equation}
\dot{\phi}_\nu = \omega_\nu + \bm{Q}_\nu' \cdot \dot{\bm{r}}_c,
\label{phi}
\end{equation}
that depends on the velocity of the wave packet. Because $H_c$ has the same translation symmetry as $H_0$, its eigenstates are still the Bloch states of $H_0$ as
\begin{equation}
\ket{\psi_{n\bm{k}}(\tilde{\bm{r}})} = e^{ik_\mu \tilde{r}^\mu} \ket{u_{n\bm{k}}(\tilde{\bm{r}})}
\end{equation} 
but with a shifted position $\tilde{{r}}^\mu = r^\mu  + \phi_\nu / Q_{\nu\mu} $ due to  $\phi_\nu (\bm{r}_c,\tau)$. 

The wave packet to the leading order can be constructed from the Bloch states as 
\begin{equation}
\ket{W_n} = \int d\bm{k} w_n(\bm{k},\tau) \ket{\psi_{n\bm{k}}(\tilde{\bm{r}})},
\end{equation}
where the electron dynamics is confined in a single band whose dispersion is $E_n(\bm{k})$ with $n$ being the band index. The interband coupling is encoded in the higher order corrections and will be studied in the next section. In the following, we denote $\bm{r}_n$ and $\bm{k}^n$ as the center of mass of the wave packet $\ket{W_n}$ in the phase space. To be self-consistent, we have 
\begin{equation}
{k}^{n}_\mu  = \bra{W_n}{k}_\mu\ket{W_n} = \int d\bm{k} |w_n(\bm{k},\tau)|^2  k_\mu
\end{equation}
that indicates $w_n(\bm{k},\tau)$ is highly localized at $\bm{k}^n$, and
\begin{widetext}
\begin{equation}
\begin{split}
{r}_n^\mu & = \bra{W_n} r^\mu \ket{W_n} = \int d\bm{k}'\int d\bm{k} w_n^*(\bm{k}',\tau)w_n(\bm{k},\tau) \bra{\psi_{n\bm{k}'}(\tilde{\bm{r}})} \left(-i\partial_{k_\mu} e^{ik_\gamma r^\gamma}\right)   e^{i k_\gamma \phi_\nu/Q_{\nu\gamma}}\ket{u_{n\bm{k}}(\tilde{\bm{r}})} \\
&=-i \int d\bm{k}'\int d\bm{k} w_n^*(\bm{k}',\tau) w_n(\bm{k},\tau) \bra{\psi_{n\bm{k}'}} \partial_{k_\mu}  \ket{\psi_{n\bm{k}}} - \phi_\nu/Q_{\nu\mu} + i\int d\bm{k}'\int d\bm{k} w_n^*(\bm{k}',\tau)w_n(\bm{k},\tau) \bra{\psi_{n\bm{k}'}}  e^{ik_\gamma\tilde{r}^\gamma}  \ket{ \partial_{k_\mu} u_{n\bm{k}}} \\
& = - \partial_{k_\mu^n} {\rm arg}[w(\bm{k}^n,\tau)] - \phi_\nu/Q_{\nu\mu} + i\braket{u_{n\bm{k}^n}| \partial_{k_\mu^n}u_{n\bm{k}^n}} 
\label{rc0}
\end{split}
\end{equation}
\end{widetext}
where we have used the integration by parts and the fact that 
\begin{equation}
\begin{split}
& \bra{\psi_{n\bm{k}'}(\tilde{\bm{r}})}  e^{ik_\gamma\tilde{r}^\gamma}  \ket{ \partial_{k_\mu} u_{n\bm{k}}(\tilde{\bm{r}})}  \\
=& \frac{1}{N}\sum_{\bm{R}} e^{i(\bm{k}-\bm{k}')\cdot\bm{R}} \frac{1}{s_0} \int_{\rm UC} d\bm{\bm{\xi}} u_{n\bm{k}'}(\bm{\xi}) e^{i(\bm{k}-\bm{k}')\cdot\bm{\xi}} \partial_{{k}_\mu} u_{n\bm{k}}(\bm{\xi})  \\
=& \delta(\bm{k}-\bm{k}') \bra{u_{n\bm{k}'}(\bm{\xi})}  e^{i(\bm{k}-\bm{k}')\cdot\bm{\xi}}   \ket{ \partial_{k_\mu} u_{n\bm{k}}(\bm{\xi})}.
\end{split}
\end{equation}
Here $N$ is the number of unit cells (UC), $s_0$ is the area of a UC, and we parameterize $\tilde{\bm{r}} =\bm{R}+\bm{\xi} $ where $\bm{R}$ is the lattice vector and $\bm{\xi}$ is within the UC centered at the the origin. Therefore $\int d\tilde{\bm{r}} = \sum_{\bm{R}} \int_{\rm UC} d\bm{\xi}$ since $u_{n\bm{k}}(\bm{R}+\bm{\xi}) = u_{n\bm{k}}(\bm{\xi})$. 

The equation of motion can be obtained from the Lagrangian 
$L =\bra{W_n} i\hbar\partial_\tau - H \ket{W_n}$.
For the first term, we have 
\begin{widetext}
\begin{equation}
\begin{split}
\bra{W_n} i\hbar\partial_\tau \ket{W_n} = & \int d\bm{k}'\int d\bm{k} w_n^*(\bm{k}',\tau) \bra{\psi_{n\bm{k}'}(\tilde{\bm{r}})} i\hbar\partial_\tau w_n(\bm{k},\tau)  e^{ik_\mu\tilde{r}^\mu}\ket{u_{n\bm{k}}(\tilde{\bm{r}})} \\
= &i\hbar \int d\bm{k} w_n^*(\bm{k},\tau) \partial_\tau w_n(\bm{k},\tau) 
- \hbar \int d\bm{k} |w_n(\bm{k},\tau)|^2 k_\mu \dot{\phi}_\nu/Q_{\nu\mu}  + 
i\hbar \int d\bm{k} |w_n(\bm{k},\tau)|^2 \braket{u_{n\bm{k}}(\tilde{\bm{r}})|\partial_\tau u_{n\bm{k}}(\tilde{\bm{r}})} \\
=& -\hbar  { \partial_{\tau} {\rm arg}[w_n(\bm{k}^n,\tau)] }
- \hbar k_\mu^n \dot{\phi}_\nu/Q_{\nu\mu}  + 
i\hbar \dot{\phi}_\nu \braket{u_{n\bm{k}^n}|\partial_{\phi_\nu} u_{n\bm{k}^n}}  \\
=& \hbar\dot{k}^n_\mu {\partial_{\bm{k}^n_\mu} {\rm arg} [w_n(\bm{k}^n,\tau)]} - \hbar k_\mu^n \dot{\phi}_\nu/Q_{\nu\mu}  + 
i\hbar \dot{\phi}_\nu \braket{u_{n\bm{k}^n}|\partial_{\phi_\nu} u_{n\bm{k}^n}} 
\end{split}
\label{dt0}
\end{equation}
\end{widetext}
where we use the fact that $-\hbar \partial_{\tau} {\rm arg}[w_n(\bm{k}^n,\tau)] = -\hbar\frac{d}{dt} {\rm arg}[w_n(\bm{k}^n,\tau)] + \hbar\dot{k}^n_\mu \partial_{k^n_\mu} {\rm arg} [w_n(\bm{k}^n,\tau)]$ and the total time derivative is dropped because it just contributes a constant to the action and does not affect the equation of motion. To the leading order, the second term of the Lagrangian is
\begin{equation}
\bra{W_n}H\ket{W_n}\approx \bra{W_n}H_c\ket{W_n} = E_n(\bm{k}^n) .
\end{equation}
Now we combine Eq. (\ref{rc0}) and Eq. (\ref{dt0}) together that yields the final expression of the Lagrangian 
\begin{equation}
\begin{split}
&L(\bm{r}_n, \dot{\bm{r}}_n, \bm{k}^n, \dot{\bm{k}}^n) = \hbar\dot{k}^n_\mu \left(-r_n^\mu- \phi_\nu/Q_{\nu\mu} + i\braket{u_{n\bm{k}^n}| \partial_{k_\mu^n}u_{n\bm{k}^n}} \right)  \\
-& \hbar k_\mu^n \dot{\phi}_\nu/Q_{\nu\mu}  + 
i\hbar \dot{\phi}_\nu \braket{u_{n\bm{k}^n}|\partial_{\phi_\nu} u_{n\bm{k}^n}}  - {E}_n(\bm{k}^n). 
\end{split}
\end{equation}
According to the Euler-Lagrange equation, the leading order semi-classical equation of the motion is
\begin{equation}
\begin{split}
&\dot{r}^\mu_n = \frac{\partial {E}_n(\bm{k}^n)}{\hbar \partial{k_\mu^n}} - \dot{k}_\gamma^n\Omega^{\mu\gamma}_{n,kk} - \dot{\phi}_\nu\Omega^{\mu\nu}_{n,k\phi}, \\
& \dot{k}_\mu^n = 0.
\end{split}
\end{equation}
where the Berry curvatures 
\begin{equation}
\Omega^{\mu\gamma}_{n,kk}  =i \left( \partial_{k^n_\mu} \braket{u_{n\bm{k}^n}| \partial_{k_\gamma^n}u_{n\bm{k}^n}} - \partial_{k^n_\gamma} \braket{u_{n\bm{k}^n}| \partial_{k_\mu^n}u_{n\bm{k}^n}} \right) , 
\end{equation}
\begin{equation}
\Omega^{\mu\nu}_{n,k\phi}  = i \left(\partial_{k^n_\mu} \braket{u_{n\bm{k}^n}| \partial_{\phi_\nu}u_{n\bm{k}^n}} - \partial_{\phi_\nu} \braket{u_{n\bm{k}^n}| \partial_{k_\mu^n}u_{n\bm{k}^n}}  \right),
\end{equation}
are respectively defined in the crystal momentum space and the hybrid momentum space consist of the crystal momentum $k^n_\mu$ and phason mode $\phi_\nu$. In this sense, $\phi_\nu$ does play the same role as the crystal momentum and the space dimension is transcendent accordingly. The dynamics of $\phi_\nu$ follows Eq. (\ref{phi}) which is part of the equation of motion. The new contribution $- \dot{\phi}_\nu\Omega^{\mu\nu}_{n,k\phi}$ to the velocity of wave packet is due to the translational motion of the periodically modulated potential.

\section{Higher order corrections to the equation of motion}\label{AB}
To include the higher order corrections to the equation of motion, we consider the wave packet with the first order corrections. According to the perturbation theory, the eigenstate of $H_c + H_1$ up to the first order is 
\begin{equation}
\begin{split}
\ket{\psi_{n\bm{k}}(\tilde{\bm{r}})} &= e^{ik_\mu\tilde{r}^\mu} \left( \ket{u_{n\bm{k}}(\tilde{\bm{r}})} + 
\sum_{m\neq n} c_m(\bm{k}) \ket{u_{m\bm{k}}(\tilde{\bm{r}})}   \right) \\
&= 
e^{ik_\mu\tilde{r}^\mu}  \left( \ket{u_{n\bm{k}}(\tilde{\bm{r}})}  + \ket{u_{n\bm{k}}^{(1)} (\tilde{\bm{r}}) }  \right),
\end{split}
\end{equation}
where $\ket{u_{n\bm{k}}^{(1)} (\tilde{\bm{r}}) }$ is the first order correction to the eigenstate and 
\begin{equation}
c_m(\bm{k}) = \frac{\bra{u_{m\bm{k}} (\tilde{\bm{r}}) }H_1\ket{u_{n\bm{k}}(\tilde{\bm{r}})}}{E_n(\bm{k})-E_m(\bm{k})}.
\end{equation}
Thus the wave packet should be constructed from the corrected eigenstates as 
\begin{equation}
\ket{W_n}=\ket{W_n^{(0)}} + \ket{W_n^{(1)}} = \int d\bm{k}w_n(\bm{k},\tau)e^{ik_\mu\tilde{r}^\mu}  \left( \ket{u_{n\bm{k}}(\tilde{\bm{r}})}  + \ket{u_{n\bm{k}}^{(1)} (\tilde{\bm{r}}) }  \right)
\end{equation}
The self-consistent conditions of the wave packet are still $k^n_\mu = \bra{W_n}k_\mu\ket{W_n}$ and
\begin{equation}
r_n^\mu = \bra{W_n^{(0)}} r^\mu\ket{W_n^{(0)}} + \bra{W_n^{(0)}} r^\mu\ket{W_n^{(1)}} + \bra{W_n^{(1)}} r^\mu\ket{W_n^{(0)}}
\label{cm}
\end{equation}
up to the first order. Here the first term is the same as Eq. (\ref{rc0}) and the second term is 
\begin{equation}
\begin{split}
\bra{W_n^{(0)}} r^\mu\ket{W_n^{(1)}}&  = i  \braket{u_{n\bm{k}^n}  | \partial_{k^n_\mu} u_{n\bm{k}^n}^{(1)}},
\end{split}
\end{equation}
where we use the fact that $\braket{u_{n\bm{k}}|u_{n\bm{k}}^{(1)}} = 0$. Therefore the center of mass is
\begin{equation}
\begin{split}
r_n^\mu &= - \partial_{k_\mu^n} {\rm arg}[w_n(\bm{k}^n,\tau)] - \phi_\nu/Q_{\nu\mu} + i\braket{u_{n\bm{k}^n}| \partial_{k_\mu^n}u_{n\bm{k}^n}} \\
& +  i  \braket{u_{n\bm{k}^n}  | \partial_{k^n_\mu} u_{n\bm{k}^n}^{(1)}} +  i  \braket{u_{n\bm{k}^n}^{(1)} | \partial_{k^n_\mu} u_{n\bm{k}^n}},
\label{rc1}
\end{split}
\end{equation}
and the first order corrections come from the interband coupling. 

For the Lagrangian $L =\bra{W_n} i\hbar\partial_\tau - H \ket{W_n}$, the first term now becomes
\begin{equation}
\begin{split}
\bra{W_n}i\hbar\partial_\tau \ket{W_n} &= \bra{W_n^{(0)}} i\hbar\partial_\tau\ket{W_n^{(0)}} + \bra{W_n^{(0)}} i\hbar\partial_\tau\ket{W_n^{(1)}} \\
&+ \bra{W_n^{(1)}} i\hbar\partial_\tau\ket{W_n^{(0)}}
\end{split}
\end{equation}
where the leading order term  $\bra{W_n^{(0)}} i\hbar\partial_\tau\ket{W_n^{(0)}}$ is the same as Eq. (\ref{dt0}) and the first order correction is 
\begin{equation}
\bra{W_n^{(0)}} i\hbar\partial_{\tau}\ket{W_n^{(1)}} = i\hbar \dot{\phi}_\nu  \braket{u_{n\bm{k}_n}  | \partial_{\phi_\nu} u_{n\bm{k}_n}^{(1)}}.
\end{equation}
The dispersion relation is corrected to the first order as
\begin{equation}
\bra{W_n}H\ket{W_n}\approx \bra{W_n}H_c+H_1\ket{W_n} =  \tilde{E}_n(\bm{k}^n) .
\end{equation}
Now the Lagrangian becomes
\begin{widetext}
\begin{equation}
\begin{split}
L(\bm{r}_n, \dot{\bm{r}}_n, \bm{k}^n, \dot{\bm{k}}^n) &= \hbar\dot{k}^n_\mu {\partial_{k^n_\mu} {\rm arg} [w_n(\bm{k}^n,\tau)]} - \hbar k_\mu^n \dot{\phi}_\nu/Q_{\nu\mu}  + 
i\hbar \dot{\phi}_\nu \braket{u_{n\bm{k}^n}|\partial_{\phi_\nu} u_{\bm{k}^n}} 
+  i\hbar \dot{\phi}_\nu  \braket{u_{n\bm{k}_n}  | \partial_{\phi_\nu} u_{n\bm{k}_n}^{(1)}}
+ i\hbar \dot{\phi}_\nu  \braket{u_{n\bm{k}_n}^{(1)}  | \partial_{\phi_\nu} u_{n\bm{k}_n}}\\
&=\hbar\dot{k}^n_\mu \left( - r_n^\mu - \phi_\nu/Q_{\nu\mu} + i\braket{u_{n\bm{k}^n}| \partial_{k_\mu^n}u_{n\bm{k}^n}}  +  i  \braket{u_{n\bm{k}^n}  | \partial_{k^n_\mu} u_{n\bm{k}^n}^{(1)}} +  i  \braket{u_{n\bm{k}^n}^{(1)} | \partial_{k^n_\mu} u_{n\bm{k}^n}} \right) - \hbar k_\mu^n \dot{\phi}_\nu/Q_{\nu\mu} \\
&\quad + 
i\hbar \dot{\phi}_\nu \braket{u_{n\bm{k}^n}|\partial_{\phi_\nu} u_{\bm{k}^n}} 
+  i\hbar \dot{\phi}_\nu  \braket{u_{n\bm{k}_n}  | \partial_{\phi_\nu} u_{n\bm{k}_n}^{(1)}}
+ i\hbar \dot{\phi}_\nu  \braket{u_{n\bm{k}_n}^{(1)}  | \partial_{\phi_\nu} u_{n\bm{k}_n}}
\end{split}
\end{equation}
\end{widetext}
according to Eq. (\ref{rc1}). The Euler-Lagrange equation yields the semi-classical equation of motion with the higher order corrections as
\begin{equation}
\begin{split}
&\dot{r}^\mu_n = \frac{\partial \tilde{E}_n(\bm{k}^n)}{\hbar \partial{k^n_\mu}} - \dot{k}_\gamma^n \tilde{\Omega}^{\mu\gamma}_{n,kk} - \dot{\phi}_\nu\tilde{\Omega}^{\mu\nu}_{n,k\phi}, \\
& \dot{k}_\mu^n = 0,
\end{split}
\end{equation}
where the Berry curvatures with the higher order corrections are 
\begin{equation}
\begin{split}
\tilde{\Omega}^{\mu\gamma}_{n,kk}  =& {\Omega}^{\mu\gamma}_{n,kk} +  i \left( \partial_{k^n_\mu} \braket{u_{n\bm{k}^n}| \partial_{k_\gamma^n}u_{n\bm{k}^n}^{(1)}} - \partial_{k^n_\gamma} \braket{u_{n\bm{k}^n}| \partial_{k_\mu^n}u_{n\bm{k}^n}^{(1)}} \right) \\
&+  i \left( \partial_{k^n_\mu} \braket{u_{n\bm{k}^n}^{(1)}| \partial_{k_\gamma^n}u_{n\bm{k}^n}} - \partial_{k^n_\gamma} \braket{u_{n\bm{k}^n}^{(1)}| \partial_{k_\mu^n}u_{n\bm{k}^n}} \right),
\end{split}
\end{equation}

\begin{equation}
\begin{split}
\tilde{\Omega}^{\mu\nu}_{n,k\phi} =& {\Omega}^{\mu\nu}_{n,k\phi} + i \left(\partial_{k^n_\mu} \braket{u_{n\bm{k}^n}| \partial_{\phi_\nu}u_{n\bm{k}^n}^{(1)}} - \partial_{\phi_\nu} \braket{u_{n\bm{k}^n}| \partial_{k_\mu^n}u_{n\bm{k}^n}^{(1)}}  \right) \\
&+ i \left(\partial_{k^n_\mu} \braket{u_{n\bm{k}^n}^{(1)}| \partial_{\phi_\nu}u_{n\bm{k}^n}} - \partial_{\phi_\nu} \braket{u_{n\bm{k}^n}^{(1)}| \partial_{k_\mu^n}u_{n\bm{k}^n}}  \right).
\end{split}
\end{equation}
Apparently, the equation of motion with the higher order corrections are still in the same form as the leading order equation of motion derived in the section above. The modified dispersion relation and Berry curvature with first order corrections are induced by the interband coupling.

\section{Unify the  dummy index}\label{AC}

In the notation above, the index $\nu=\{1,2,\cdots\}$ is for the ancillary dimensions
associated with phason modes of the translational motion of the periodic modulation, while $\mu,\gamma = \{x,y\}$ are for the physical dimensions. To unify the dummy indices, we introduce the hybrid momentum space  spanned by the generalized momentum 
\begin{equation}
\tilde{\bm{k}} = (k_x, k_y, \phi_1,\phi_2, \cdots).
\end{equation}
In the current notation, the equation of motion becomes 
\begin{equation}
\dot{r}^\mu_n = \frac{\partial \tilde{E}_n(\tilde{\bm{k}})}{\hbar \partial{\tilde{k}_\mu}} - \dot{\tilde{k}}_\nu\tilde{\Omega}^{\mu\nu}_{n}, 
\label{em}
\end{equation}
where the superscript of $\tilde{\bm{k}}^n$ is dropped hereafter and in the main text for simplification. Now $\mu,\nu,\gamma = \{ x,y,1,2,\cdots \}$ are equivalent. However,  Eq. (\ref{em}) is meaningful only for $\mu=x$ or $y$ since the center of mass $\bm{r}_n$ is confined in the real space. Here the unified Berry curvature is
\begin{equation}
\begin{split}
&\tilde{\Omega}^{\mu\nu}_{n} = \Omega_n^{\mu\nu} + \Omega_n^{(1)\mu\nu},  \\
&\Omega_n^{\mu\nu}  = i \left( \partial_{\tilde{k}_\mu} \braket{u_{n\tilde{\bm{k}}}| \partial_{\tilde{k}_\nu}u_{n\tilde{\bm{k}}}} - \partial_{\tilde{k}_\nu} \braket{u_{n\tilde{\bm{k}}}| \partial_{\tilde{k}_\mu}u_{n\tilde{\bm{k}}}} \right), \\
&\Omega_n^{(1)\mu\nu} = i \left( \partial_{\tilde{k}_\mu} \braket{u_{n\tilde{\bm{k}}}| \partial_{\tilde{k}_\nu}u_{n\tilde{\bm{k}}}^{(1)}} - \partial_{\tilde{k}_\nu} \braket{u_{n\tilde{\bm{k}}}| \partial_{\tilde{k}_\mu}u_{n\tilde{\bm{k}}}^{(1)}} \right) \\
&\;\qquad +  i \left( \partial_{\tilde{k}_\mu} \braket{u_{n\tilde{\bm{k}}}^{(1)}| \partial_{\tilde{k}_\nu}u_{n\tilde{\bm{k}}}} - \partial_{\tilde{k}_\nu} \braket{u_{n\tilde{\bm{k}}}^{(1)}| \partial_{\tilde{k}_\mu}u_{n\tilde{\bm{k}}}} \right),
\end{split}
\end{equation}
where $\Omega_n^{\mu\nu}$ is the leading order Berry curvature and $\Omega_n^{(1)\mu\nu}$ is the first order correction to the Berry curvature. The dynamics of the generalized momentum follows
\begin{equation}
\dot{\tilde{k}}_\nu =
\begin{cases}
&0, \;\; {\rm for}\;\; \nu =x,y \\
&\omega_\nu  + {Q}_{\nu\gamma}' \dot{r}^\gamma_n , \;\; {\rm for}\;\; \nu =1,2,\cdots
\end{cases}.
\label{dk}
\end{equation}
The deformation of the periodic modulation potential couples $\dot{\tilde{k}}_\nu$ with $\dot{r}_n^\gamma$ that results in the nonlinear response of the system to the deformation. The nonlinear response function will be derived in the next section.

\section{Nonlinear response function}\label{AD}

The current density due to the motion and deformation of periodic modulation potential (induced by spin textures) is
\begin{equation}
j^\mu = e \sum_n \int d^2 k \dot{r}^\mu_n f(E_n-E_F) D_n(\bm{r},\tilde{\bm{k}})  
\label{current}
\end{equation}
where $D_n(\bm{r},\tilde{\bm{k}})$ is the phase space density of states  (PSDOS). In the absence of deformation, the phase space is homogeneous  and the PSDOS is a constant as $D_n=1/(2\pi)^d$ where $d$ is the dimension of the system. However, due to the deformation, the phase space is inhomogeneous and we need to consider the modification of PSDOS. The basic idea is to study the time evolution of phase space volume element $\Delta V = \Delta \bm{r} \Delta \tilde{\bm{k}}$ that follows $(1/\Delta V)d\Delta V/d t = \nabla_{\bm{r}} \cdot \dot{\bm{r}} +  \nabla_{\tilde{\bm{k}}} \cdot \dot{\tilde{\bm{k}}} $ \cite{Xiao2005}. The equation can be solved \cite{Price_Four_2015} and yields the modified PSDOS
\begin{equation}
\begin{split}
D_n(\bm{r},\tilde{\bm{k}}) = \frac{1}{\Delta V} = &\frac{1}{(2\pi)^2} \bigg(1+\frac{1}{2} Q_{\mu\nu}' \tilde{\Omega}_n^{\nu\mu} + \\
&\frac{1}{64} \varepsilon^{\mu\nu\gamma\delta} Q_{\mu\nu}'Q_{\gamma\delta}'\varepsilon_{\eta\rho\alpha\beta} \tilde{\Omega}_n^{\eta\rho} \tilde{\Omega}_n^{\alpha\beta} \bigg),
\label{PSDOS}
\end{split}
\end{equation} 
where $\varepsilon^{\mu\nu\gamma\delta}$ and $\varepsilon_{\eta\rho\alpha\beta}$ are Levi-Civita tensor with $\varepsilon^{xy12}=\varepsilon_{xy12}=1$.  Plugging Eq. (\ref{dk}) into Eq. (\ref{em}) and retaining the terms up to the second order in $\omega_\nu$ and $Q_{\nu\mu}'$ yields 
\begin{equation}
\dot{r}^{\mu }_{n}=\frac{\partial \tilde{E}_n  }{\hbar \partial k_{\mu }}-\omega _{\nu }  \tilde{\Omega}_n^{\mu \nu }-\left(\frac{\partial \tilde{E}_n}{\hbar \partial k_{\gamma }}-\omega _{\delta }  \tilde{\Omega}_n^{\gamma \delta }-\frac{\partial \tilde{E}_n }{\hbar \partial k_{\eta}}Q_{ \delta\eta}' \tilde{\Omega}_n^{\gamma \delta }\right)Q_{\nu \gamma }' \tilde{\Omega}_n^{\mu \nu },
\label{dr}
\end{equation}
which is in the same form of Eq. (\ref{r2}) but with modified dispersion relation and Berry curvature due to the first order corrections. In the following, we will show that our finally results are not changed by the first order corrections which  can be dropped. Now we substitute Eq. (\ref{PSDOS}) and Eq. (\ref{dr}) into Eq. (\ref{current}) and retain the terms up the second order  in $\omega_\nu$ and $Q_{\nu\mu}'$ that gives 
\begin{widetext}
\begin{equation}
\begin{split} 
j^\mu= &e\sum_n \int \frac{d^2k}{(2\pi)^2}f(\tilde{E}_n-E_F) 
\Bigg[
\left(\frac{\partial \tilde{E}_n  }{\hbar \partial k_{\mu }}-\omega _{\nu }  \tilde{\Omega}_n^{\mu \nu}\right) 
+\left( \omega _{\delta }  \Omega_n^{\gamma \delta }Q_{\nu \gamma }' \Omega_n^{\mu \nu} - \frac{1}{2}\omega _{\nu }  \Omega_n^{\mu \nu}Q_{\gamma\delta}' \Omega_n^{\delta\gamma} \right) - 
\left( \frac{\partial \tilde{E}_n}{\hbar \partial k_{\gamma }}Q_{\nu \gamma }' \tilde{\Omega}_n^{\mu \nu} - \frac{1}{2}\frac{\partial \tilde{E}_n}{\hbar \partial k_{\mu }}Q_{\gamma\delta}' \tilde{\Omega}_n^{\delta\gamma}\right) \\
&+\Bigg(\frac{\partial {E}_n }{\hbar \partial k_{\eta}}Q_{ \delta\eta}' \Omega_n^{\gamma \delta }
Q_{\nu \gamma }' \Omega_n^{\mu \nu} - \frac{1}{2}\frac{\partial {E}_n}{\hbar \partial k_{\gamma }} Q_{\nu \gamma }' \Omega_n^{\mu \nu}Q_{\eta\delta}' \Omega_n^{\delta\eta}
+  \frac{1}{64} \frac{\partial {E}_n  }{\hbar \partial k_{\mu }} \varepsilon^{\lambda\nu\gamma\delta} Q_{\lambda\nu}'Q_{\gamma\delta}'\varepsilon_{\eta\rho\alpha\beta} \Omega_n^{\eta\rho} \Omega_n^{\alpha\beta} \Bigg)
\Bigg].
\end{split}
\label{jmu}
\end{equation}
\end{widetext}
Here we collect different terms with similar orders and forms into four groups enclosed by brackets.  The integral of the first group just gives Eq. (\ref{j0}) where higher order corrections are dropped. Since we focus on the Fermi energy in the bulk energy gap that remains open upon weak perturbation, the modified group velocity $\partial_{k_\mu} \tilde{E}_n/\hbar$ and the first order correction to the Berry curvature $\Omega_n^{(1)\mu\nu}$ vanish under the integration over hybrid Brillouin zone \cite{gao2014}. Therefore, the transported charge described by Eq. (\ref{q0})  is unaffected by the higher order corrections. Notably, the second group in Eq. (\ref{jmu}) is unaffected by the higher order corrections. Because $\Omega_n^{\mu\nu}=-\Omega_n^{\nu\mu}$, $Q_{\mu\nu}'=-Q_{\nu\mu}'$ , and the dummy indices can be freely replaced, we have 
\begin{equation}
\begin{split}
&\omega _{\delta }  \Omega_n^{\gamma \delta }Q_{\nu \gamma }' \Omega_n^{\mu \nu} - \frac{1}{2}\omega _{\nu }  \Omega_n^{\mu \nu}Q_{\gamma\delta}' \Omega_n^{\delta\gamma} \\
=& \frac{\omega_\nu Q_{\gamma\delta}'}{2}\left( \Omega_n^{\mu \nu } \Omega_n^{\gamma \delta}+\Omega_n^{\mu \gamma } \Omega_n^{\delta \nu}+\Omega_n^{\mu \delta } \Omega_n^{\nu\gamma } \right)
\end{split}
\end{equation}
for the second group. Here we use the transformation 
$\omega _{\delta }  \Omega_n^{\gamma \delta }Q_{\nu \gamma }' \Omega_n^{\mu \nu} \rightarrow 
\omega _{\nu }  \Omega_n^{\nu\gamma}Q_{\gamma\delta}' \Omega_n^{\mu \delta}$ under $\delta \leftrightarrow \nu$ and 
$\omega _{\delta }  \Omega_n^{\gamma \delta }Q_{\nu \gamma }' \Omega_n^{\mu \nu} \rightarrow 
\omega _{\nu }  \Omega_n^{\delta\nu}Q_{\gamma\delta}' \Omega_n^{\mu \gamma}$ under $\gamma\leftrightarrow \nu$ and then $\nu \leftrightarrow \delta$. Therefore, the integral of the first two groups yields the nonlinear response function Eq. (\ref{j}). 

Next, we are going to prove the integral of the last two groups are zero. For the third group, we use the same trick as above and we get 
\begin{equation}
\begin{split}
&\frac{\partial \tilde{E}_n}{\hbar \partial k_{\gamma }}Q_{\nu \gamma }' \tilde{\Omega}_n^{\mu \nu} - \frac{1}{2}\frac{\partial \tilde{E}_n}{\hbar \partial k_{\mu }}Q_{\gamma\delta}' \tilde{\Omega}_n^{\delta\gamma} \\
= &\frac{Q_{\gamma\delta}'}{2\hbar}\left( 
\frac{\partial \tilde{E}_n}{\partial k_\mu}\tilde{\Omega}^{\gamma\delta}_n
+\frac{\partial \tilde{E}_n}{\partial k_\gamma}\tilde{\Omega}^{\delta\mu}_n
+\frac{\partial \tilde{E}_n}{\partial k_\delta}\tilde{\Omega}^{\mu\gamma}_n \right)
\label{third}.
\end{split}
\end{equation}
For a fully occupied band at zero temperature, the integral of Eq. (\ref{third}) yields
\begin{equation}
 \frac{Q_{\gamma\delta}'}{2\hbar} \int d^2k \tilde{E}_n \left(
 \frac{\partial \tilde{\Omega}^{\gamma\delta}_n}{\partial k_\mu}
+\frac{\partial \tilde{\Omega}^{\delta\mu}_n}{\partial k_\gamma}
+\frac{\partial \tilde{\Omega}^{\mu\gamma}_n}{\partial k_\delta}
\right),
\end{equation}
where we use the integration by parts and the fact that both $\tilde{E}_n$ and $\tilde{\Omega}^{\mu\nu}_n$ are periodic over the Brillouin zone. The three terms in the bracket equal to zero according to the Bianchi identity \cite{weinberg}.  Therefore, the third group in Eq. (\ref{jmu}) does not contribute to the current density. The last group 
vanishes due to the antisymmetric property of $\Omega_n^{\mu\nu}$ and $Q_{\mu\nu}'$. To show the three terms are indeed canceled out, we take $j^y$ as an example and we collect all the terms associated with $\frac{\partial E_n}{\hbar\partial k_y}Q_{1y}'Q_{2x}'$. Here the first two terms give 
\begin{equation}
\frac{\partial E_n}{\hbar\partial k_y}Q_{1y}'Q_{2x}'\left( \Omega_n^{x1} \Omega_n^{y2} - \Omega_n^{y1}\Omega_n^{x2} \right)
\label{t1}
\end{equation}
and the last term yields 
\begin{equation}
\frac{\partial E_n}{\hbar\partial k_y}Q_{1y}'Q_{2x}'\left(  - \Omega_n^{x1} \Omega_n^{y2} + \Omega_n^{y1}\Omega_n^{x2}  \right)
\label{t2}
\end{equation}
where we use the property $\Omega_n^{\mu\nu}Q_{\gamma\delta}'= \Omega_n^{\nu\mu}Q_{\delta\gamma}'$, $\Omega_n^{\mu\nu}\Omega_n^{\gamma\delta}= \Omega_n^{\nu\mu}\Omega_n^{\delta\gamma}$, and $Q_{\mu\nu}' Q_{\gamma\delta}'= Q_{\nu\mu}' Q_{\delta\gamma}'$. Obviously, Eq. (\ref{t1}) and Eq. (\ref{t2}) are opposite to each other, and the last group has no contribution to the current density.

\section{Monte Carlo simulation}\label{AF}
We present details of the model and numerical simulations for the spin textures. 
We here consider two models in order to describe three multiple-$Q$ states: one is the effective spin model derived from the Kondo lattice model for the double-$Q$ and triple-$Q$ collinear spin textures and the other is the frustrated spin model for the triple-$Q$ skyrmion spin texture. 

The effective spin model is given by~\cite{Hayami2017,Takagi2018} 
\begin{align}
\label{eq:KLM}
&\mathcal{H}=  2\sum_\nu
\left[ -\tilde{J}\left\{\alpha (S^x_{\bm{Q_{\nu}}} S^x_{-\bm{Q_{\nu}}}+S^y_{\bm{Q_{\nu}}} S^y_{-\bm{Q_{\nu}}})+S^z_{\bm{Q_{\nu}}} S^z_{-\bm{Q_{\nu}}}\right\} \right. \nonumber \\ 
 &\left.+\tilde{K} \left\{\alpha (S^x_{\bm{Q_{\nu}}} S^x_{-\bm{Q_{\nu}}}+S^y_{\bm{Q_{\nu}}} S^y_{-\bm{Q_{\nu}}})+S^z_{\bm{Q_{\nu}}} S^z_{-\bm{Q_{\nu}}}\right\}^2 \right]
- A \sum_{i} (S_i^z)^2, 
\end{align}
where $\bm{S}_{\bm{q}}$ is the Fourier transform of $\bm{S}_i$ and $\bm{Q}_\nu$ [$\nu=1,2$ ($1,2,3$) for the square (triangular) lattice] are the equivalent wave vectors for the multiple peaks in the bare susceptibility of itinerant electrons. 
The bilinear interaction $\tilde{J}=1$ represents the effective Ruderman-Kittel-Kasuya-Yosida interaction and the positive biquadratic interaction $K=N \tilde{K}$ {with $N$ the number of sites} represents the four-spin interactions leading to the multiple-$Q$ instability. 
$\alpha$ is introduced as the anisotropy parameter for exchange coupling where $\alpha=0$ represents the Ising-type coupling.  
The third term represents the easy-axis anisotropy ($A>0$) for localized spins. 

The simulations for the effective spin model are carried out with the classical Monte Carlo simulations at low temperatures. 
Our simulations are carried out with the standard Metropolis local updates. 
The results are obtained for systems with $N=96\times 96$ sites under periodic boundary conditions.  
In each simulation, we perform simulated annealing to find the low-energy spin configuration by gradually reducing the temperature with the rate $T_{n+1} = a T_n$, where $T_n$ is the temperature in the $n$th step. We set the initial temperature {$T_0 = 1.0-10.0$}. We take ${a} = 0.99995$-$0.99999$ and the final temperature is typically taken at $T = 0.001$-$0.01$. 
{We also start the simulations from the multiple-$Q$ spin configurations, such as the double-$Q$ and triple-$Q$ collinear spin textures.
We determine the magnetic phase by comparing their energies of the obtained magnetic patterns in simulations.
}
The double-$Q$ collinear spin texture in Fig.~\ref{fig1}(a) is obtained at $T=0.01$, $K=0.5$, $\alpha=1.0$, $A=0.5$, and $|\bm{Q}|=2\pi/6$ on the square lattice, while the triple-$Q$ collinear spin texture in Fig.~\ref{fig1}(b) is obtained at $T=0.001$, $K=1.0$, $\alpha=0.0$, $A=0.0$, and $|\bm{Q}|=2\pi/6$ on the triangular lattice. 

The frustrated spin model on the triangular lattice is given by~\cite{Hayami2016} 
\begin{align}
\label{eq:Ham}
\mathcal{H}= \sum_{\langle i,j \rangle} J_{ij} \mathbf{S}_i \cdot \mathbf{S}_j - H \sum_{i}  S^z_i - A \sum_i (S^z_i)^2, 
\end{align} 
where we consider the ferromagnetic nearest-neighbor coupling $J_1=-1$ and antiferromagnetic third nearest-neighbor coupling $J_3=0.5$ in the first term, which gives the ordering vector $|\bm{Q}|=2\pi/5$. 
The second and third terms represent the Zeeman coupling to an external magnetic field and the easy-axis anisotropy, respectively. 

The optimal magnetic phases in the frustrated spin model are obtained from the Monte Carlo simulations based on the Metropolis algorithm. 
The lattices have $N=100\times 100$ spins and the periodic boundary conditions. In the simulations, the $10^5-10^7$ Monte Carlo sweeps measurements are performed after equilibration. 
The triple-$Q$ skyrmion spin texture in Fig.~\ref{fig1}(c) is obtained at $T=0.01$, $H=0.27$, and $A=0.5$.

\section{Time-dependent current}\label{AG}
In the lattice model, the current through a bond {connecting} lattice sites $i$ and $j$ is 
\begin{equation}
I_{ij}(\tau) = -i\frac{e}{\hbar} \sum_{E_n\le E_F,\sigma} \bra{\psi_n(\tau)} c_{i,\sigma}^\dagger t_{ij} c_{j,\sigma} -c_{j,\sigma}^\dagger t_{ji} c_{i,\sigma} \ket{\psi_n(\tau)},
\end{equation} 
where $\sigma=\uparrow$ or $\downarrow$ for spin degree of freedom and $\ket{\psi_n(\tau)}$ is the time evolution of the eigenstate $\ket{\psi_n}$ that is obtained by exact diagonalization of the Hamiltonian Eq. (\ref{H}) at $\tau=0$. The time evolution of eigenstates follows the time-dependent Schr{\"o}dinger equation  $i\hbar{\partial_\tau} \ket{\psi_n(\tau)} = \mathcal{H}(\tau) \ket{\psi_n(\tau)}$,
which is solved numerically by the Runge-Kutta method. The total current $I_x$ ($I_y$) is the sum of all the bond currents through a cross section perpendicular to the $x$ ($y$) direction.

\section{Anomalous charge transport for the spin texture ansatz}\label{AE}

{To compare with the numerical results for the spin textures from Monte Carlo simulation, we study the anomalous charge transport for the spin texture ansatz described by }
\begin{equation}\label{S}
\bm{S_i} = \left(0,0,\sum_{\nu}\cos ( \bm{Q}_\nu \cdot \bm{r}_i + \phi_\nu) \right).
\end{equation}
{The spin texture ansatz is exactly periodic and hence well quantized anomalous charge transport is expected.}

\begin{figure}[t]
  \begin{center}
  \includegraphics[width=8.5 cm]{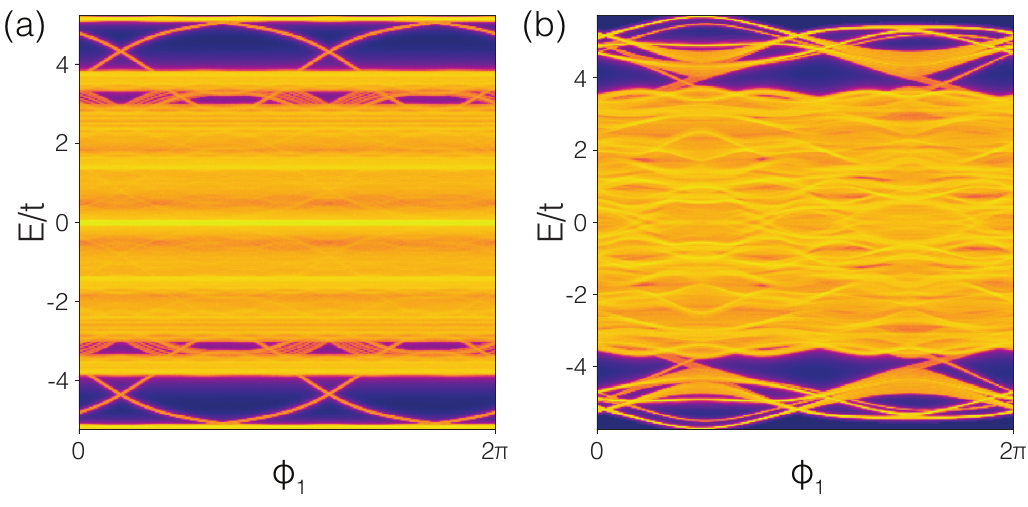}
  \end{center}
\caption{Electronic spectral functions of the magnets with the double-$Q$ collinear spin texture (a) and triple-$Q$ collinear spin texture (b) depicted by the ansatz as a function of $\phi_1$. Here the purple (yellow) denotes low (high) density of states.   } 
  \label{figs1}
\end{figure}

Here we consider a  double-$Q$ collinear spin texture with $\bm{Q}_1=(2\pi/a,0)$ and $\bm{Q}_2=(0,2\pi/a)$, and a triple-$Q$ collinear spin texture with $\bm{Q}_1=(-2\pi/a,2\sqrt{3}\pi/a)$, $\bm{Q}_2=(-2\pi/a,-2\sqrt{3}\pi/a)$, and $\bm{Q}_3=(4\pi/a,0)$.
We first study the electronic spectrum to identify the topological property of the system. For conduction electrons hopping on a square lattice (whose lattice constant is set to unity), we fix $a=5$ for both the double-$Q$ and triple-$Q$ spin textures depicted by Eq. (\ref{S}). While for the triple-$Q$ spin texture, to make it commensurate with the lattice, we compress the lattice constant along the $y$ direction to $1/\sqrt{3}$ such that each supercell also contains $5\times 5$ lattice sites. The energy spectra as a function of $\phi_1$ for the double-$Q$ and triple-$Q$ spin textures when $J=1.5t$ and $B=0$ in Eq. (\ref{H}) are shown in Figs. \ref{figs1}(a) and \ref{figs1}(b), respectively.  Here we use the open boundary condition along the $x$ direction and the periodic boundary condition along the $y$ direction. Apparently, there are topological edge states in the bulk energy gaps indicating the system is topologically nontrivial. To characterize the nontrivial topology, we calculate the Chern numbers of the system. Here we focus on the Fermi energy in the bottom bulk energy gaps in Fig. \ref{figs1}. In the 4D hybrid momentum space of the double-$Q$  spin texture, there are six first Chern numbers:  {$C_1^{x1}=C_1^{y2}=2$ and $C_1^{xy}=C_1^{x2}=C_1^{y1}=C_1^{12}=0$, and one second Chern number: $C_2^{x1y2}=-2$}. In the 5D hybrid momentum space of the triple-$Q$ spin texture, there are ten first Chern numbers: {$C_1^{x1}=C_1^{x2}=-C_1^{y1}=C_1^{y2}=-2$ and $C_1^{xy}=C_1^{x3}=C_1^{y3}=C_1^{12}=C_1^{13}=C_1^{23}=0$, and five second Chern numbers: $C_2^{x1y2}=-2$ and $C_2^{xy13}=C_2^{xy23}=C_2^{x123}=C_2^{y123}=0$}. The first Chern numbers on different 2D planes and second Chern numbers on different 4D hypersurfaces are independent of the other momenta because the bulk energy gaps are open in the entire hybrid momentum space.

\begin{figure}[t]
  \begin{center}
  \includegraphics[width=8.5 cm]{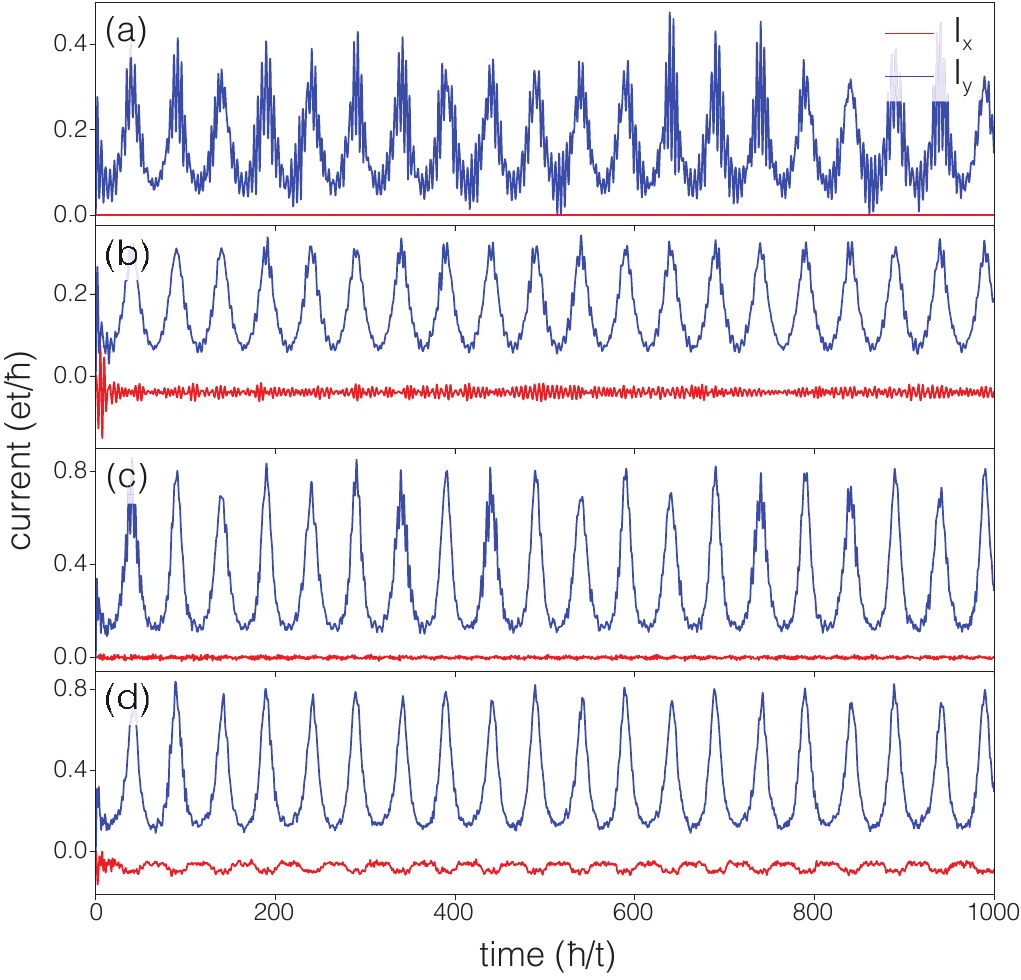}
  \end{center}
\caption{Currents along the $x$ and $y$ directions for the magnets with multiple-$Q$ collinear spin textures depicted by the ansatz. (a) and (b) Currents for the double-$Q$ spin texture without  and with deformation, respectively. (c) and (d) Currents for the  triple-$Q$ spin texture without and with deformation, respectively.} 
  \label{figs2}
\end{figure}

{We then study the anomalous charge transport induced by the translational motion of the spin textures, which is parameterized  by $\phi_\nu = \omega_\nu\tau$. Here we consider the spin texture moving in the $y$ direction same as that in the main text.  For the double-$Q$ spin texture, we have $\omega_1=0$ and $\omega_2=2\pi t/100\hbar$, while for the triple-$Q$ spin texture, we have $\omega_1=-\omega_2=2\pi t/100\hbar$ and $\omega_3=0$. Thus the period of motion of both the double-$Q$ and triple-$Q$ spin textures is $\Delta\tau=100\hbar/t$. We calculate the currents due to the motion of spin textures on a $40\times40$ lattice (containing $8\times8$ supercells).
For the double-$Q$ spin texture, the currents over ten periods are shown in \ref{figs2}(a).  
By integrating the currents over time, the average transported charges in one period are quantized as $ q^x=0.00e$ and $q^y=15.99e$. According to Eq. (\ref{q0}), the transported charges are $q^x=eN^x {\omega}_2 C_1^{x2}/|{\omega}_2|=0e$ and $q^y=eN^y {\omega}_2 C_1^{y2}/|{\omega}_2|=16e$. 
For the triple-$Q$ spin texture, the currents over ten periods are shown in Fig. \ref{figs2}(c) that gives $q^x=0.00e$ and $q^y=31.97e$. The transported charges are $q^x=eN^x {\omega}_1 C_1^{x1}/|{\omega}_1|+eN^x {\omega}_2 C_1^{x2}/|{\omega}_2|=0e$ and $q^y=eN^y {\omega}_1 C_1^{y1}/|{\omega}_1|+eN^y {\omega}_2 C_1^{y2}/|{\omega}_2|=32e$ according to Eq. (\ref{q0}). Apparently, the numerical results show a very good agreement with Eq. (\ref{q0}).
}

To trigger the nonlinear response, one needs to deform the spin texture. For the double-$Q$ spin texture, we apply a shear deformation of the spin texture by deforming the $\bm{Q}_1$ vector to $\bm{Q}_1=(2\pi/5,2\pi/20)$. By shifting the spin texture in the same way mentioned above, we   obtain $q^x=-4.00e$ and $q^y=16.00e$ from the currents in Fig. \ref{figs2}(b). The numerical results are also consistent with $q^x=eL^xC_2^{x21y}{\omega}_2{Q}_{1y}'/2\pi|{\omega}_2|=-4e$ and $q^y=16e$ according to Eq. (\ref{q}). For the triple-$Q$ spin texture, we apply a shear deformation of the spin texture by deforming the $\bm{Q}_\nu$ vectors to $\bm{Q}_1=(-2\pi/5,2\sqrt{3}\pi/5-2\sqrt{3}\pi/20)$, $\bm{Q}_2=(-2\pi/5,-2\sqrt{3}\pi/5-2\sqrt{3}\pi/20)$, and $\bm{Q}_3=(4\pi/5,2\sqrt{3}\pi/10)$.  In this case, the currents over ten periods are shown in Fig. \ref{figs2}(d) that gives $q^x=-7.98e$ and $q^y=31.96e$. Eq. (\ref{q0}) yields $q^x=eL^xC_2^{x12y}{\omega}_1{Q}_{2y}'/2\pi|{\omega}_1|+eL^xC_2^{x21y}{\omega}_2{Q}_{1y}'/2\pi|{\omega}_2|=-8e$ and $q^y=32e$ as before that agree with the numerical results. The numerical results of anomalous charge transport for the spin texture ansatz are perfectly quantized and exhibit very good agreement with the response function.

\bibliography{references}

\end{document}